\documentclass[twocolumn,prb, floatfix, showpacs, superscriptaddress]{revtex4}
\usepackage[utf8]{inputenc}
\usepackage[T1]{fontenc}
\usepackage{graphicx}
\usepackage{amsmath}
\usepackage{latexsym}
\usepackage{amssymb}
\usepackage{hyperref}
\usepackage{pstricks}
\usepackage{array}
\usepackage{soul}
\newcommand{\ket}[1]{\left| #1 \right>} 
\newcommand{\up}{\uparrow}
\newcommand{\down}{\downarrow}

\def\be{\begin{equation}}
\def\ee{\end{equation}}

\begin{document}

\title{Theoretical, numerical, and experimental study of a flying qubit electronic interferometer}
\author{Tobias Bautze}
\affiliation{Univ. Grenoble Alpes, Inst. NEEL, F-38042 Grenoble, France}
\affiliation{CNRS, Inst. NEEL, F-38042 Grenoble, France}
\author{Christoph S\"ussmeier}
\affiliation{Univ. Grenoble Alpes, Inst. NEEL, F-38042 Grenoble, France}
\affiliation{CNRS, Inst. NEEL, F-38042 Grenoble, France}
\author{Shintaro Takada}
\affiliation{Department of Applied Physics, University of Tokyo, Bunkyo-ku, Tokyo, 113-8656, Japan}
\author{Christoph Groth}
\affiliation{CEA-INAC/UJF Grenoble 1, SPSMS UMR-E 9001, F-38054 Grenoble, France}
\author{Tristan Meunier}
\affiliation{Univ. Grenoble Alpes, Inst. NEEL, F-38042 Grenoble, France}
\affiliation{CNRS, Inst. NEEL, F-38042 Grenoble, France}
\author{Michihisa Yamamoto}
\affiliation{Department of Applied Physics, University of Tokyo, Bunkyo-ku, Tokyo, 113-8656, Japan}
\affiliation{PRESTO, JST, Kawaguchi-shi, Saitama, 332-0012, Japan}
\author{Seigo Tarucha}
\affiliation{Department of Applied Physics, University of Tokyo, Bunkyo-ku, Tokyo, 113-8656, Japan}
\affiliation{Center for Emergent Matter Science (CEMS), RIKEN, Wako, Saitama 351-0198, Japan}
\author{Xavier Waintal}
\affiliation{CEA-INAC/UJF Grenoble 1, SPSMS UMR-E 9001, F-38054 Grenoble, France}
\author{Christopher B\"auerle}
\affiliation{Univ. Grenoble Alpes, Inst. NEEL, F-38042 Grenoble, France}
\affiliation{CNRS, Inst. NEEL, F-38042 Grenoble, France}
\date{\today}

\begin{abstract}

We discuss an electronic interferometer recently measured by Yamamoto et al. This  "flying quantum bit'' experiment showed  quantum oscillations between electronic trajectories of two tunnel-coupled wires connected via an Aharanov-Bohm ring. We present a simple scattering model as well as a numerical microscopic  model to describe this experiment. In addition, we present new experimental data to which we confront our numerical results. While our analytical model provides basic concepts for designing the flying qubit device, we find that our numerical simulations allow to reproduce detailed features of the transport measurements such as in-phase and anti-phase oscillations of the two output currents as well as a smooth phase shift when sweeping a side gate. Furthermore, we find remarkable resemblance for the magneto conductance oscillations in both conductance and visibility between simulations and experiments within a specific parameter range. 

\end{abstract}
\maketitle

\section{Introduction}

In quantum information science, solid state approaches are attractive as they are easily scalable.
The coherent transport of information in such systems is usually coded into the degrees of freedom of the electron, either spin \cite{delft-esr} or charge \cite{hayashi,petta_10}.
The advantage of the spin degree of freedom lies in the fact that it is well protected from the electrostatic environment whereas the charge degree of freedom is easily measurable \cite{field}.
Over the last decade, a variety of interesting electronic devices have been proposed and tested, such as Fabry-Perot  \cite{camino_05,willet_pnas_09,halperin_11} or Mach-Zehnder interferometers in the quantum Hall regime \cite{heiblum_03, samuelsson_04,neder_07,preden_08}.
In particular, these quantum Hall systems are particularly appealing as they could be operated as flying qubits, where the quantum information can be manipulated during flight, due to the absence of backscattering.
The fact that now an individual electron charge can be controlled and manipulated in such systems, opens the possibility to perform electron quantum optics experiments at the single electron level \cite{feve_ses_07,hermelin_nature_11,mc-nail_nature_11,feve_science-ex_13,glattli_nature_13}.
In a recent experiment by Yamamoto et al. \cite{michi_nnano_12}, the charge degree of freedom has been exploited to gain full electrical control of a flying qubit.
The system is composed of an Aharonov-Bohm (AB) ring connected to two tunnel-coupled wires at each side, which can act as beam splitters for the ballistic electrons.
In this case, the system behaves like a Mach-Zehnder interferometer for ballistic electrons \cite{heiblum_03,preden_08}.
Compared to quantum Hall systems, this system is more easily scalable and no magnetic field is in principle necessary to operate the device.
It is made from a semiconductor heterostructure and electrostatic surface gates define the borders of the interferometer by locally depleting the two-dimensional electron gas while the two middle gates allow to tune the tunnel-coupling of the two wires at the entrance and exit of the AB ring (see Fig.\ \ref{fig:sample}) \cite{wieck_apl,d-coupler}.
In the original experiment, several interesting experimental features have been observed.
By applying a relatively small negative gate voltage on the right tunnel gate $V_{t}$ and connecting the left tunnel gate to ground, only the middle island will be depleted, but essentially no tunnel barrier will be formed in the tunnel-coupled wire region.
In this case the system behaves as a two-terminal device and the currents $I_{u}$ and $I_{d}$ are identical. When sweeping the gate voltage more negative, the tunnel barrier between the upper and lower channel can be tuned.
For a sufficiently negative voltage the \emph{in-phase} oscillations of $I_{u}$ and $I_{d}$ turn surprisingly into \emph{anti-phase} oscillations.
Other interesting observations have been made such as the possibility to control the partition between the two output currents by using the tunnel-coupling gates.
It has also been shown that this interferometer does not suffer from backscattered electrons which encircle the AB loop and hence allows to perform reliable phase shift measurements \cite{shintaro_13}.
In this article, we address all the experimental findings of ref.\cite{michi_nnano_12}  by means of a simplified theoretical model that can account for several features observed in the experiments. In order to capture the more subtle features we perform numerical simulations and confront them with the experiment. Our simulations show that the majority of the experimentally observed features can be well explained within the Landauer-B\"uttiker scattering formalism \cite{landauer_IBM,landauer_PM,buttiker_multi-channel,buttiker_4-terminal}.
\section{Summary of the main experimental features}
\begin{figure}[htbp]
	\includegraphics[width=7cm]{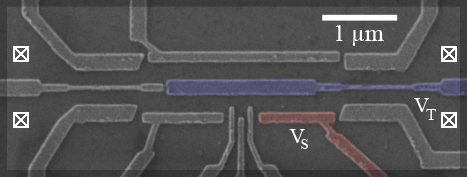}
	\caption{\label{fig:sample} Scanning electron microscope image of the flying qubit sample. The outer metallic gates (light grey) define the borders of the interferometer while the right tunnel gate (blue) allows for depletion of the centre AB island as well as for adjustment of the tunnel barrier of the split-wire. Changing the voltage of the side gate (red) allows to induce a phase shift in the lower branch (see text).  White squares represent the ohmic contacts.  }
\end{figure}

The flying qubit sample is tailored within a two-dimensional electron gas (2DEG) of density $n_s= 3.2 \cdot 10^{11} \mathrm{cm^{-2}}$ and a mobility $\mu=0.86 \cdot10^6 \mathrm{cm^2/Vs}$ made from a GaAs/AlGaAs heterostructure by metallic (Ti/Au) surface gates (see Fig.\ \ref{fig:sample}).
The 2DEG is situated 90 nm below the surface, hence the electrostatic potential applied to the surface gates leads to a smooth potential change in the 2DEG over roughly the same distance.
For the measurements, a bias voltage is applied to the lower left contact (the upper left contact is floating) and the current is measured simultaneously in the right upper and lower contact.
To observe Aharonov-Bohm oscillations, the magnetic field is swept over a field range of approximately $\pm$\,80 mT. Above this field one suffers from Shubnikov-de-Haas oscillations.
As the right tunnel gate voltage $V_T$ is switched on, the sample goes through a series of different regimes. Initially, the Aharonov-Bohm loop does not exist and one only observes universal conductance fluctuations.
As $V_T$ is swept more negative, the AB region gets more depleted, yet the gate does not affect much the tunnel-coupled region. We refer to this regime as the Strong Coupling Regime (SCR).
Upon further increasing negative $V_T$, one enters the regime of main interest:
the Weak Coupling Regime (WCR) where the 2DEG is also partly depleted underneath the tunnel gate. In this regime, there is a finite coupling between the up and down channels in the wire region. Finally, upon further increase of $V_T$, one enters a regime where the upper and lower channels are decoupled.
Below we list the different experimental features when either scanning the tunnel gate through the different regimes or by scanning the side gate voltage of $V_s$ (see Fig.\ \ref{fig:sample}) and which we attempt to reproduce with analytical as well as numerical approaches.
\begin{table}[h!]
  \begin{tabular}{ ll }
    \multicolumn{2}{l}{\ }\\
    $\bullet$ (P1a)\  & \ \textbf{Magnetic field sweep in SCR:} \\
                      & \ \ \ $I_{u}$ and $I_{d}$ show almost identical  in-phase\\
                      & \ \ \ oscillations with magnetic field \\
    $\bullet$ (P1b)\  & \ \textbf{Magnetic field sweep in WCR} \\
    							    & \ \ \ $I_{u}$ and $I_{d}$ show anti-phase oscillations \\
  	  					  	  & \ \ \ with magnetic field \\
    $\bullet$ (P2a)\  & \ \textbf{Side gate sweep in SCR:} \\
    					  			& \ \ \ AB oscillations show phase jumps for \\
  	  			  				& \ \ \ $I_{u}(B)$ and $I_{d}(B)$  \\
    $\bullet$ (P2b)\  & \ \textbf{Side gate sweep in WCR:}  \\
    			  					& \ \ \ AB oscillations show a smooth phase \\
  	  	  						& \ \ \ shift for $I_{u}$ and $I_{d}$  \\
    $\bullet$ \ (P3)\ & \ \textbf{Anti-phase oscillations in WCR:}\\
                      & \ \ \ $I_{u}$ and $I_{d}$ show oscillations with respect \\
                      & \ \ \ to the tunnel gate voltage $V_T$\\
   
  \end{tabular}
\end{table}

\section{A minimum (scattering) theory of the flying qubit.}

\begin{figure}[htbp]
	\includegraphics[width=8.5cm]{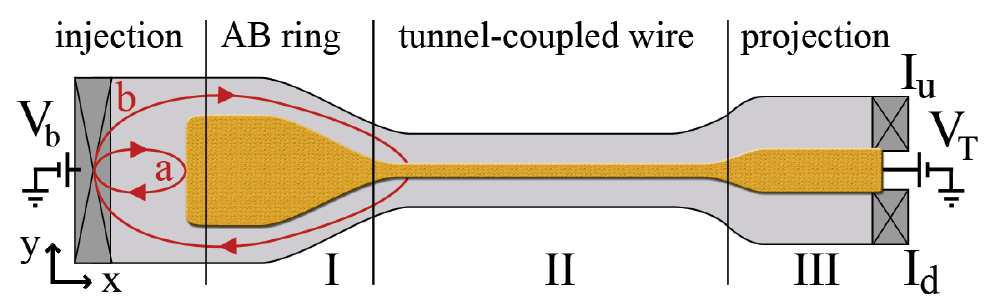}
	\caption{\label{fig:cartoon} Schematic of the system used for the modelling. We divide the sample into several regions: the injection region on the left, the central Aharonov-Bohm region  and the tunnel-coupled wire on the right. The tunnel-coupled wire can further be split into three regions I, II and III:  the two wires are coupled only in region II and decoupled upon entering in region I or III.  a), b) correspond to two possible backscattered electron trajectories that could, in principle, contribute to the reflection amplitude. }
\end{figure}

In this section we develop a minimum scattering approach which captures the main features of the experiment. In Fig.\ \ref{fig:cartoon} we show a sketch of the actual device used in Ref.\ \onlinecite{michi_nnano_12}. 
Here, the two left contacts of the original device have been replaced by a single contact. 
This is a simplification which has no effect on the experimentally observed results. 
As we show below, the flying qubit can be manipulated even in such a three-terminal configuration \cite{wieck_apl,d-coupler}.
The difference from the four-terminal device only appears as a shift of the AB oscillation phase by $\pi$/2.

The device consists of several distinct regions: the injecting region on the left, the central Aharonov-Bohm region and the tunnel-coupled wire on the right which are modelled by their respective scattering matrices, $S_{inj}$, $S_{ab}$ and $S_{tw}$.
In this section, we assume for simplicity that a single channel, labeled \textbf{u}p and \textbf{d}own, is propagating inside each arm of the interferometer (this assumption will be relaxed in the numerics performed on the microscopic model).

A general property of a 2DEG is the smoothness of the electrostatic potential generated by the surface gates felt by the electrons in the 2DEG as it is situated approximately 90 nm below the surface.
Here we assume that the scattering is mostly forward, which is valid in a specific configuration. Suppose the width of each wire is kept unchanged between the ring and coupled-wire region, the potential change $\Delta V$  of each wire at this transition region is simply defined by the tunnel coupling. For small $\Delta V$ (smooth potential change), the length scale of the tunnelling $\hbar \,v_x / \Delta V $ becomes much larger than that of the potential change at the transition region, which suppresses tunnel-backscattering into the other arm. Since the potential is also smooth with respect to the Fermi wave length, intra-wire backscattering is also suppressed. 

\subsection{Generalities}
Let us first discuss the structure of the different scattering matrices for the three distinct regions. $S_{inj}$ is rather arbitrary. It is characterised by the amplitude for an electron injected from the left to be transmitted into the upper $a_u$ and lower $a_d$ channel (which, we assume, does not depend on the injecting channel).
The probability $|S_{bs}|^2$ for an electron to be backscattered into the injecting electrode is obtained from the current conservation law $|S_{bs}|^2+|a_u|^2+|a_d|^2=1$.
The Aharonov-Bohm region is translation invariant along $x$, and therefore, no backscattering occurs there and the upper/lower electrons are simply transmitted into their respective arms.
The down transmission amplitude picks up an Aharonov-Bohm phase $e^{i 2\pi \Phi/\Phi_0}$ with respect to the upper one where $\Phi_0=h/e$ is the flux quantum and $\Phi= S B$ is the magnetic flux through the Aharonov-Bohm region ($B$: magnetic field, $S$: effective surface of the AB ring).
Next we describe the tunnel-coupled wire region.
If the variation of the lateral confinement potential at the transition region between the AB ring and the coupled wire is smooth, the electrons can only be transmitted and ring encircling paths due to backscattering can be ignored. Once the electrons are scattered into the tunnel-coupled wire, the corresponding transmission matrix can be parametrised as
\be
t_{tw}=
\left(\begin{array}{cc} t_{uu} & t_{ud} \\ t_{du} & t_{dd} \end{array} \right)
\ee
with $t_{tw}t_{tw}^\dagger=1$ (current conservation).
Summing up all the probability amplitudes for the different paths taken by the electrons, we arrive at a Mach-Zehnder like expression
\be
t_{u}=  t_{uu} a_u +  e^{i 2\pi \Phi/\Phi_0} t_{ud} a_d
\ee
\be
t_{d}=  t_{du} a_u +  e^{i 2\pi \Phi/\Phi_0} t_{dd} a_d
\ee
where $t_{u/d}$ is the total amplitude for an electron injected from the left to be transmitted in the upper/lower right electrode. The corresponding currents are given by the Landauer formula\cite{landauer_IBM,landauer_PM},
\be
\frac{dI_{u/d}}{dV_b}=\frac{2e^2}{h}  T_{u/d}.
\ee
Here $T_{u/d}=|t_{u/d}|^2$ is the total transmission probability from the left to the right upper/lower electrode.

Before going further, let us discuss the conclusions that can be already drawn at this stage. First, properties (P1a) and (P2a) are rather natural: in the strong coupling limit, the system is essentially a {\it two-terminal}
Aharonov-Bohm device where forward scattering in the wire makes the upper and lower current homogeneous (P1a).
Onsager symmetry imposes that $G(B)=G(-B)$ which leads to phase rigidity \cite{buttiker_phase_rigidity_1,buttiker_phase_rigidity_2} as the phase of the conductance can only take multiples of $\pi$ at zero magnetic field, as observed in many experiments \cite{benoit-webb_prl86,benoit+hasselbach_prl_97}.
Current-conservation imposes that the injected current $I_{inj}$ = $I_{bs} + I_u + I_d$, hence property (P1b) simply translates into $I_{bs}$ being {\it independent} on $B$.
From the sample geometry we can also make some assumptions about the electron trajectories that dominantly contribute to the transport properties.
Figure \ref{fig:cartoon} shows two different backscattered trajectories that could potentially
contribute to the electronic transport. In particular, in order to observe an Aharonov-Bohm effect on $I_{bs}$, one needs both trajectories.
We have argued, however, that trajectory b) can be neglected due to the smoothness of the confining potential that prevents backscattering at the interface between the Aharonov-Bohm and tunnel wire region, hence only a) contributes to the reflection amplitude and $I_{bs}$ is essentially independent of the magnetic field (P1b).
Property (P2b) is also due to the absence of the backscattering trajectory b). It straightforwardly leads to a realization of the two-slit experiment once the electron is injected into the AB ring. 

\subsection{Scattering matrix of the tunnel-coupled wire}

Let us now focus on the tunnelling region and compute the transmission matrix $t_{tw}$.
As seen from Fig.\ \ref{fig:cartoon}, at the entrance and exit of tunnel-coupled wire (region I and III), the tunnel barrier is infinitively high and, as a consequence, the two separated wires are fully decoupled, whereas in the central region (region II), the coupling is finite.
We suppose that the transition between the regions happens smoothly.

The eigenstate in region II for mode $\alpha$ takes the form,
\be
\Psi_\alpha (x,y)=\Psi_\alpha (y) e^{ik_\alpha x}
\ee
and we consider the situation where only two modes can propagate in the wire, hereafter labeled  the symmetric $\ket {S_{II}}$ and anti-symmetric $\ket {A_{II}}$ mode. Indeed, $\Psi_\alpha (y)$ corresponds to the solution of the 1D Schr\"odinger equation along the transverse direction $y$,
\begin{equation}
[-\frac{\hbar^2}{2m}\frac{\partial^2}{\partial y^2} +V(y)]\Psi_\alpha (y)=E \Psi_\alpha (y)
\label{s-equation}
\end{equation}
with $E= E_F -\frac{\hbar^2}{2m}k_\alpha^2$ and its two solutions in the absence of the tunnelling gate are respectively a symmetric and anti-symmetric function of $y$.
The actual wave function of these two eigenstates of the tunnel-coupled region are displayed in Fig.\ \ref{fig:S+AS}. The discussion of the evolution of these states when changing the tunnel barrier height is postponed to the next subsection where we treat this issue semi-analytically.

Upon going from region I to II, $\ket {A_{II}}$ is essentially unaffected (the weight of $\Psi_{A_{II}} (y)$
in this tunnelling region is very small so the gate hardly affects this mode) while $\ket{S_{I}}$ is smoothly transformed into $\ket {S_{II}}$ whose wave function is essentially $\Psi_{S_{I}} (y)=|\Psi_{A_{I}} (y)|$.
The two modes $\ket {S_I}$ and $\ket {A_I}$ are degenerate and can also be rewritten as combinations of the
modes that propagate in the upper ($\ket{\up}$) or lower ($\ket{\down}$) parts:
\begin{equation}
\ket{S_I}=\ket{\up}+\ket{\down} ; \ket{A_I}=\ket{\up}-\ket{\down}
\end{equation}
 The transmission matrix $t_{tw}$ can now be obtained by the following adiabatic argument: let us start with an electron in mode $\ket{\up}=(\ket{S_I}+\ket{A_I})/2$. In the beginning of region II, the wave function has smoothly evolved into $(\ket{S_{II}}+\ket{A_{II}})/2$. Towards the end of region II, the wave function has picked up mode-dependent phases and reads $(e^{ik_S L}\ket{S_{II}}+e^{ik_A L}\ket{A_{II}})/2$ where $L$ is the total length of region II. Then, the wave function is smoothly transformed into $(e^{ik_S L}\ket{S_{III}}+e^{ik_A L}\ket{A_{III}})/2$ which can be re-expressed as $(e^{ik_S L} [\ket{\up}+\ket{\down}]+
e^{ik_A L} [\ket{\up}-\ket{\down}])/2$. We can directly read from this expression the amplitude to be transmitted in the up ($(e^{ik_S L}+e^{ik_A L})/2$) and down ($(e^{ik_S L}-e^{ik_A L})/2$) channel. Repeating the procedure for  spin down we arrive at,
\be
t_{tw}=
\exp(i\frac{k_S+k_A}{2}L) \left(\begin{array}{cc} \cos(\frac{k_A-k_S}{2}L)& i\sin(\frac{k_S-k_A}{2}L) \\  i\sin(\frac{k_S-k_A}{2}L) & \cos(\frac{k_A-k_S}{2}L) \end{array} \right)
\ee

Putting everything together, the Landauer formula finally provides
\begin{align}
\label{eq:simple}
\frac{dI_{u/d}}{dV_b} &=\frac{2e^2}{h}
\big[
\frac{|a_u|^2 + |a_d|^2}{2}
\pm \frac{|a_u|^2 - |a_d|^2}{2} \cos [(k_A-k_S)L]  \nonumber \\
& \pm |a_u a_d| \sin [(k_A-k_S)L] \cos \left( 2\pi \Phi/\Phi_0 +\phi \right) \big]
\end{align}


Equation \ref{eq:simple} now provides a general analytical description of the interferometer.
Changing the amplitudes $a_u$ and $a_d$ allows to control the symmetry of the injected wave function.
When injecting into one arm only, the last term of equation \ref{eq:simple}  cancels and the system reduces to a simple split-wire.
This simple analysis shows that the currents in the upper and lower branches oscillate anti-phase as a function of $\Delta k=k_A-k_S$. Varying $\Delta k$ is equivalent to changing the tunnel gate voltage and hence explains the experimentally observed oscillations with respect to $V_T$ [property (P3)]. In a similar way, for a given $\Delta k$ the two output currents have opposite sign and will also oscillate anti-phase as a function of magnetic field (P1b).


\subsection{Semi-analytical analysis}

\begin{figure}[htbp]
	\includegraphics[width=7.5cm]{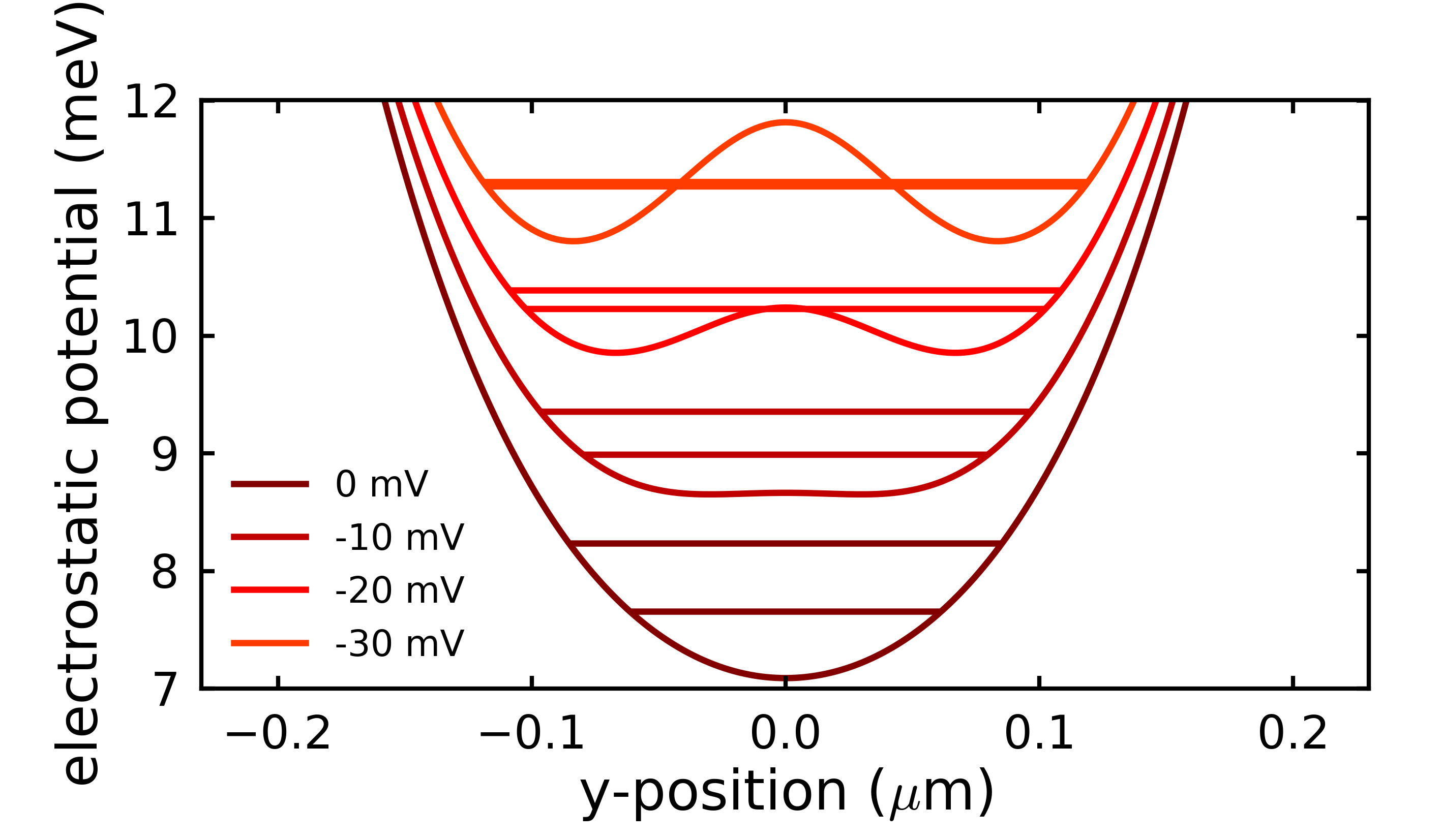}
	\caption{Electrostatic potential $V(y)$ created by the electrostatic gates defining the split-wire for different tunnel gate voltages. The horizontal lines correspond to the quantized energies of the symmetric and anti-symmetric state for each tunnel gate voltage.}
	\label{fig:wire-potential}
\end{figure}

In this subsection we would like to get some more physical insight into the experimental system by calculating the precise dependence of $\Delta k$ on the tunnel gate voltage $V_T$. This can be done by numerically solving the Schr\"odinger equation (Eq.\ \ref{s-equation}) of our system \cite{ref-quant}.  
For this we first implement the electrostatic potential felt by the electrons which are at a depth of 90 nm below the surface gates by following the approach of Ref.\ \onlinecite{sukhorukov_apl_95}, where the electrostatic potential created by a polygon surface gate is calculated by solving the Laplace equation (screening effects by the electrons in the 2DEG are however not taken into account).
The obtained electrostatic potential profile of the split-wire along the y-direction for different tunnel gate voltages is shown in Fig.\ \ref{fig:wire-potential}. 
It resembles qualitatively the one of the experimental situation of the data we present later on and has been used to realize the numerical simulation in section IV.
As can be seen from Fig.\ \ref{fig:wire-potential}  we explore in detail the crossover region between the SCR and the WCR regime.

\begin{figure}[htbp]
	\includegraphics[width=8.5cm]{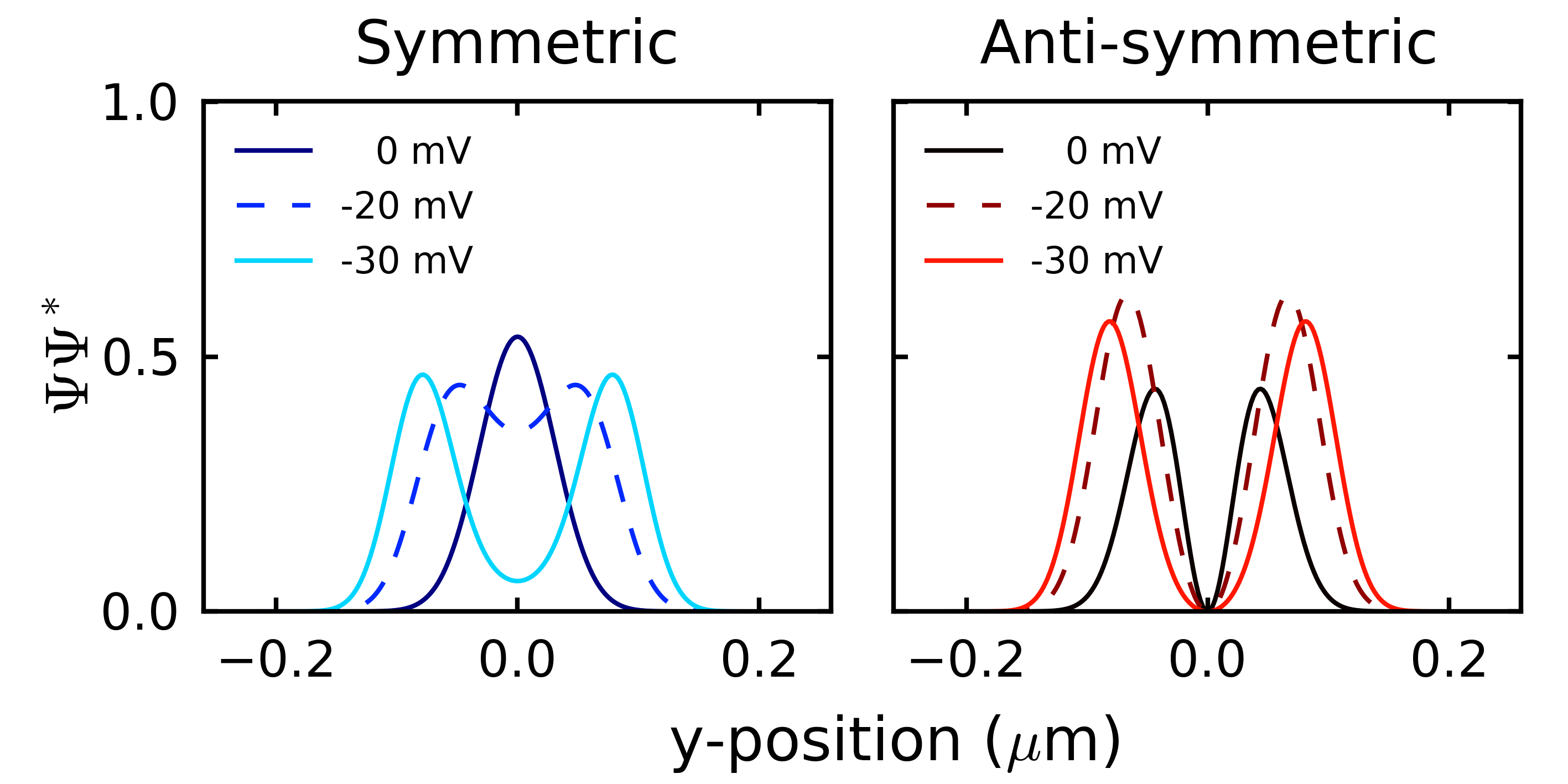}
	\caption{Normalized wave function probability of the symmetric (left) and anti-symmetric (right) state in the tunnel-coupled wire for different tunnel gate voltages.}
\label{fig:S+AS}
\end{figure}

Assuming an infinitely long tunnel-coupled wire we can then calculate numerically the actual wave function of the symmetric and anti-symmetric state as displayed in Fig.\ \ref{fig:S+AS}.
At zero tunnel gate voltage the weight of the symmetric state is pinned in the centre of the split-wire, whereas the anti-symmetric wave function has its weight within each tunnel-coupled wire. When increasing the tunnel barrier, the symmetric and the anti-symmetric wave functions are displaced differently. However, upon further increasing the tunnel barrier, the symmetric and anti-symmetric wave function become similar and finally degenerate when completely decoupled.

\begin{figure}[htbp]
	\includegraphics[width=8cm]{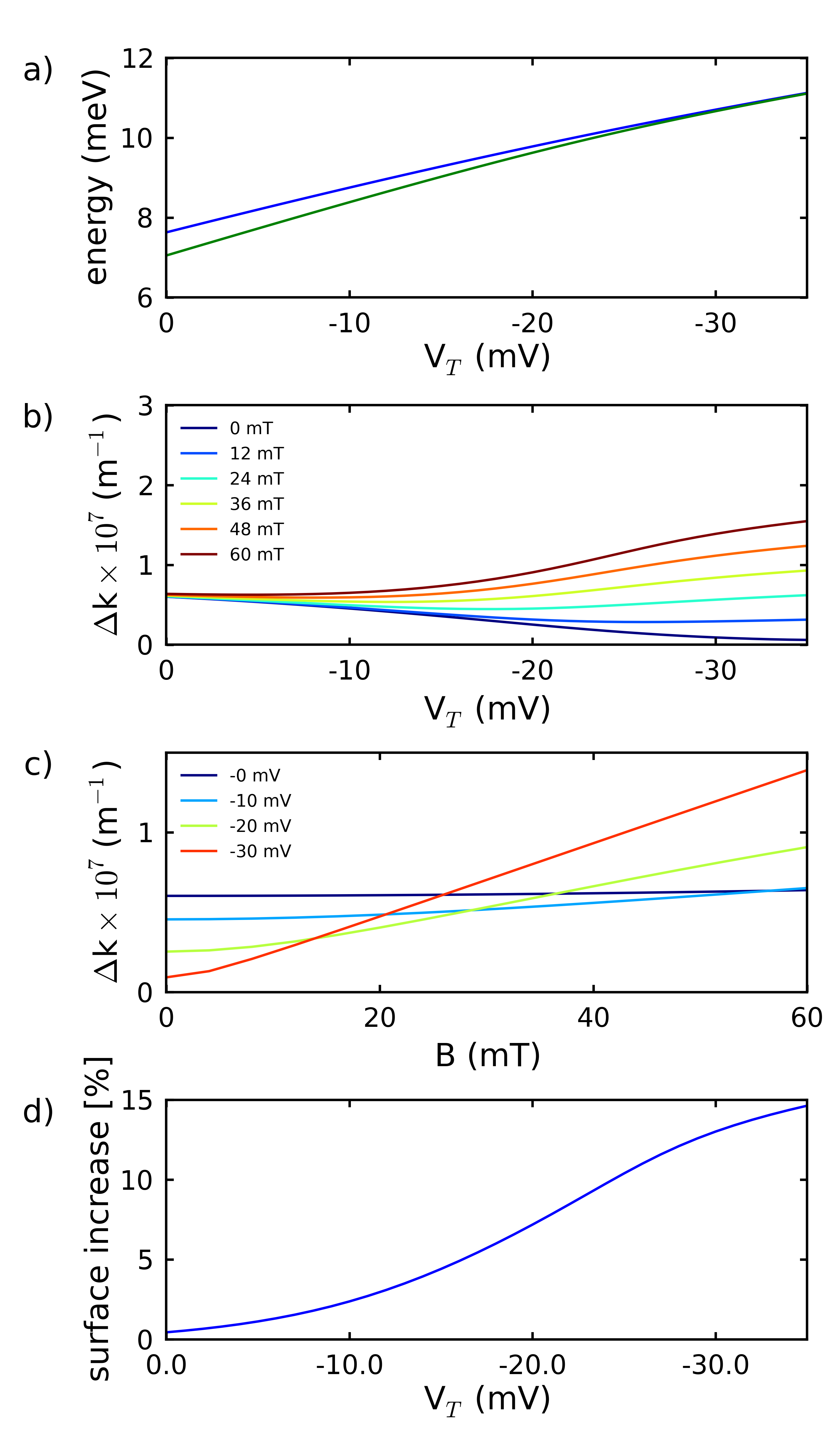}
	\caption{ a) energy dependence of the symmetric (green) and anti-symmetric (blue) state as a function of tunnel gate voltage for $B=$ 0 T. b) tunnel gate voltage dependence of $\Delta k$ for different magnetic fields. c) magnetic field dependence of $\Delta k$ for different tunnel gate voltages. d) surface area increase with respect to the AB ring calculated using the results of c). }
	\label{fig:bonding+antibonding}
\end{figure}

This can easily be seen when plotting the corresponding eigenenergies of the symmetric and anti-symmetric state as illustrated in panel a) of Fig.\ \ref{fig:bonding+antibonding}.
At zero tunnel barrier height the energy difference is simply the energy separation of the two lowest energy states of the potential created by the approximately harmonic confinement of the outer electrostatic gates of the split-wire, whereas at large tunnel barrier height, the energy difference vanishes as the two minima of the split-wire potential are decoupled.
The absolute energy values of the two states move to higher energy as the confinement potential is stronger due to the strong influence of the tunnel barrier and eventually cross the Fermi energy (in our case 11.4 meV, see section IV). 

Similarly, the dispersion relation of our system as well as the values for $\Delta k$ = $k_A-k_S$ as a function of tunnel gate voltage and magnetic field can be evaluated, which is detailed in the appendix.
At low energy, which is of interest here, we observe that the energy bands for the symmetric and anti-symmetric states are affected by the tunnel coupling as well as the magnetic field.
For zero magnetic field, $\Delta$k is decreasing when increasing the negative tunnel gate voltage (Fig.\ \ref{fig:bonding+antibonding}b). This is expected since the symmetric and anti-symmetric state become degenerate. 
The contrary is observed when applying a magnetic field. The influence of the magnetic field is to displace the wave functions with respect to the centre of the tunnel-coupled wire (see appendix). As a consequence, $\Delta$k for a given tunnel gate voltage is increasing with magnetic field. 
This can also be seen in the field dependence of $\Delta$k for fixed tunnel gate voltages (Fig.\ \ref{fig:bonding+antibonding}c). The stronger the tunnel barrier, the stronger is the increase in $\Delta$k. For a completely decoupled wire the slope of $\Delta$k with respect to magnetic field finally saturates.
As a consequence, the electrons will pick up an additional phase difference when traversing the tunnel-coupled wire. This will eventually lead to a change in oscillation period of the magneto conductance oscillations. We will come back to this point in Section IV.

\begin{figure}[htbp]
	\includegraphics[width=8cm]{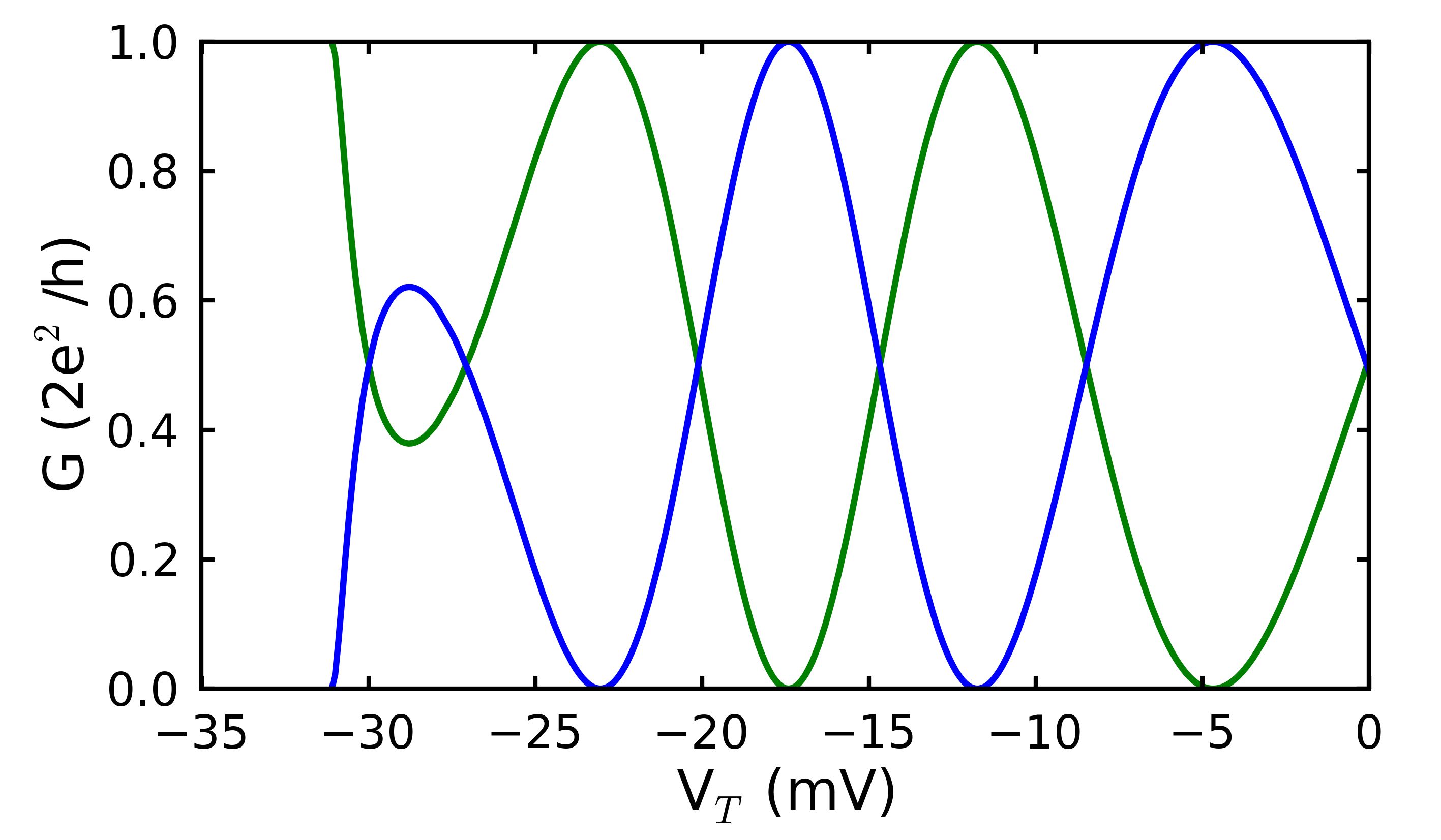}
	\caption{Conductance of the upper/lower channel as a function of tunnel gate voltage as given by Eq.(\ref{eq:simple}). The mapping between $V_T$ and $\Delta k$ was performed numerically.}
	\label{fig:VT}
\end{figure}

Having numerically determined the values of $\Delta k$  for different tunnel barrier heights, we can then compute the current in the upper (lower) branch using equation \ref{eq:simple}.
The corresponding conductance versus $V_T$ trace is shown in Fig.\ref{fig:VT}.
We clearly see that the two output currents oscillate in anti-phase with respect to the tunnel gate voltage $V_T$ as observed in the experiment.
At zero tunnel gate voltage the two output currents are equal since the upper and lower channels are strongly coupled.
Increasing the negative tunnel gate voltage induces anti-phase oscillations until the tunnel gate completely separates  the two channels (P3).
This demonstrates that the tuning of the tunnel gate allows to reach a fully electrical control of the repartition of  the output currents  into the upper/lower branch.
When the two output currents are equal, the tunnel-coupled wire behaves like a perfect beam splitter.

  \section{Microscopic theory: model and simulation}

In the preceding section we have been able to understand the underlying physics of the Aharonov-Bohm interferometer coupled to a tunnel-coupled wire by means of a simplified analytical model (complemented with a numerical calculation of the mapping between $V_T$ and $\Delta k=k_A-k_S$).
Even though the analytical model provides basic concepts for designing the flying qubit device, it relies on the assumption that encircling paths induced by backscattering are fully suppressed. In the following, we make use of a detailed microscopic model to confirm that we can indeed suppress the encircling paths for the weak tunnel coupling by correctly choosing the device configuration. We show that the main experimental features (P1)-(P3) are very well reproduced with the simulations. 
Neglecting screening and Coulomb interactions, our potential calculations do not allow to discuss precise quantitative agreement between experiments and simulations. Interestingly however, we find that for a certain parameter range, both the conductance and the visibility of the oscillation can be tuned close to what is observed in the experiments. 


\subsection{Microscopic model}

In the following numerical simulations, the sample is modelled by a simple single-band Schr\"odinger equation that includes the confining potential $V(x,y)$ due to the gate structure as well as an uniform magnetic field B.
\begin{equation}
\frac{1}{2m}[   i\hbar \vec\nabla - e\vec A(x,y)]^2 \psi(x,y) + V(x,y)\psi (x,y) = E\psi(x,y)
\label{eq:sql}
\end{equation}
For the actual simulations, the model is discretized on a square lattice with lattice constant $a$ and
we introduce the wave function $\psi_{n_x,n_y}\equiv \psi (n_x a,n_y a)$.  The discretized Schr\"odinger equation reads,
\begin{align}
   &-e^{-i\phi n_y}\psi_{n_x+1,n_y}- e^{+i\phi n_y}\psi_{n_x-1,n_y} -
\psi_{n_x,n_y+1}-\psi_{n_x,n_y-1} \nonumber \\
&+ V_{n_x,n_y} \psi_{n_x,n_y}= (E/\gamma+2) \psi_{n_x,n_y}
\end{align}
where $\gamma=\hbar^2/(2ma^2)$ and $\phi=eBa^2/\hbar$.
The numerical calculations of the differential conductances are performed with the Kwant code\cite{kwant_groth}. Kwant is a general purpose
library designed for quantum transport.
The calculations are done within the standard framework of the Landauer-B\"uttiker approach\cite{Blanter2000} which is
also equivalent, in this context, to the non-equilibrium Green's function formalism (NEGF) \cite{Wingreen-Meir_1992}.

\begin{figure}[htbp]
	\includegraphics[width=8cm]{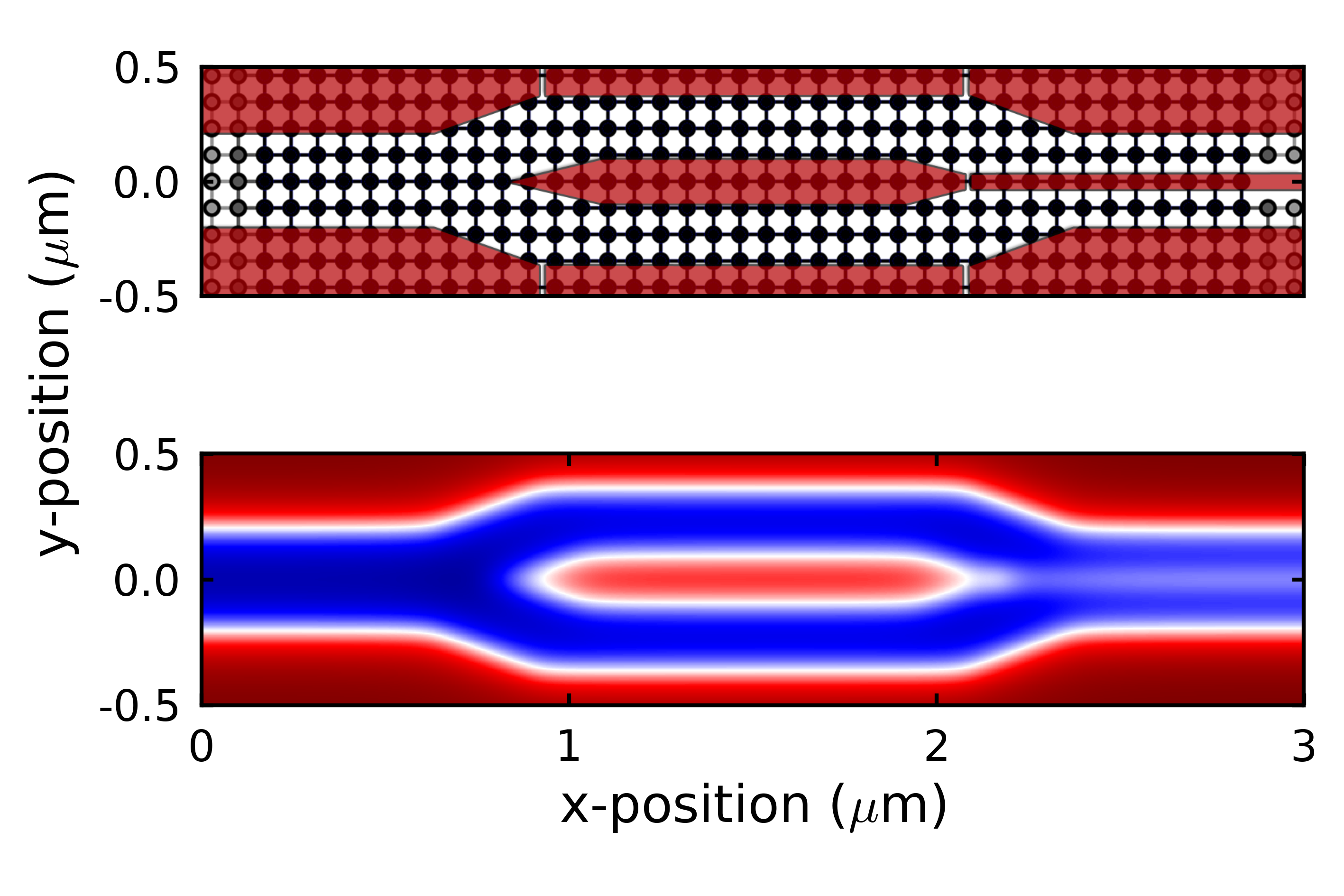}
	\caption{Top: lattice site model of the sample (see Fig.1). For clarity we only display few lattice sites. In the actual sample the lattice grid was much finer e.g. 800 $\times$ 100 lattice sites. Bottom: electrostatic potential felt by the electrons in the 2DEG, about 90 nm below the surface. }
	\label{fig:sites}
\end{figure}

The system that we used for the simulations is composed of approximately 800 $\times$ 100 lattice sites as shown in Fig.\ \ref{fig:sites}a.
We have taken a sufficiently large number of lattice sites such that no influence of the discretisation on the transport properties is observed and we can hence safely assume that we are in the continuum limit. The separation of the tunnel-coupled wire on the right side has been implemented by a smooth transition towards the ohmic contacts.  These contacts are mimicked by semi-infinite wires following the standard approach in NEGF.
In order to convert the tight-binding parameters into \emph{experimental} units \cite{waintal_KNIT},
we fix the Fermi energy of the system to 11.4 meV corresponding to an electron density of $n_s$= 3.2$\cdot$ $10^{11}$ cm$^{-2}$.
Using a lattice constant of  $a\,=\,5\,$nm we define gates and distances in real space units.

In order to provide a realistic electrostatic potential associated with the different electrostatic gates of the sample,
we follow the approach of Ref.\ \onlinecite{sukhorukov_apl_95} as briefly mentioned in section II.
The two-dimensional potential at a depth of 90 nm below the surface obtained with this approach is plotted in Fig.\ \ref{fig:sites}b.
For convenience, we also separate the right tunnel-coupled wire from the middle island, such that we can sweep its voltage independently.
This allows us to investigate the influence of the tunnel-coupled wire on the AB oscillation frequency without affecting the depleted AB area.
When performing the simulations, we adjust the gate voltages for the simulations by the following procedure:
First, we fix the energy to match the  Fermi energy and then sweep simultaneously all the outer gates to obtain the desired conductance similar to the one of the experiment. Afterwards, we sweep the desired parameter ($V_T, V_S$ or $B$) and record the two output currents.

\subsection{Comparison between numerics and experiment}

\begin{figure}[htbp]
	\includegraphics[width=8cm]{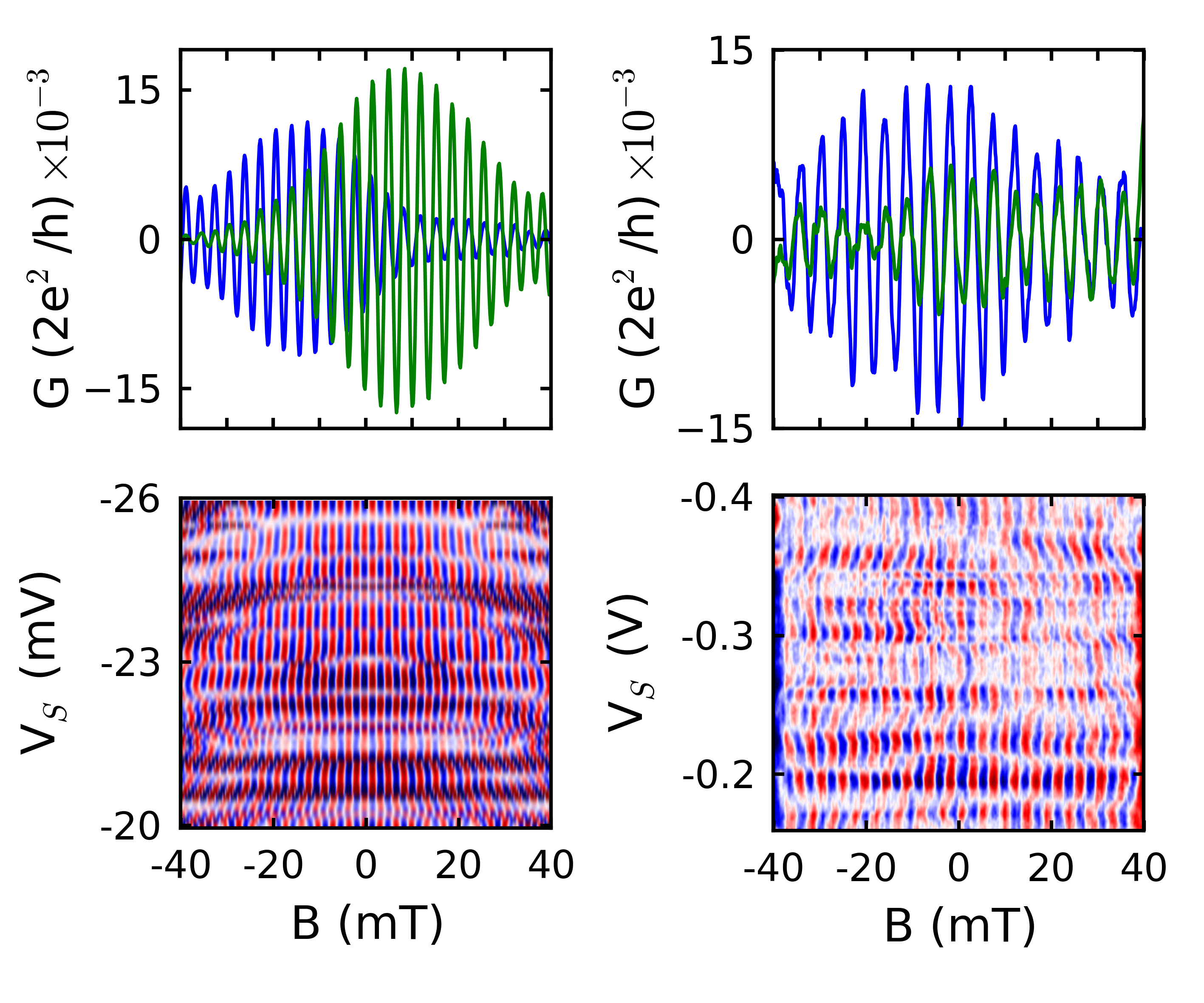}
	\caption{Magneto conductance oscillations in the strong coupling regime after subtraction of a smooth background. Left: simulations, right: experimental data. Top: magneto conductance oscillations for small tunnel gate voltage $V_T$. The blue (green) curve corresponds to the current in the upper (lower) contact. Bottom: 2D colour plot of the magneto conductance oscillations of the total transmitted current (upper - lower) as a function of side gate voltage $V_s$. Phase jumps are clearly observed in the simulated as well as experimental data. }
	\label{fig:in-phase}
\end{figure}

In the following, we first address the issue of the magneto oscillations in the strong coupling regime (P1a \& P2a).
We apply a finite gate voltage to the centre island in order to form an Aharanov-Bohm ring
and then sweep the magnetic field as well as the side gate voltage $V_{S}$. The simulated data is confronted with the experimental data in Fig.\ \ref{fig:in-phase}.
For all simulations we set the total conductance (transmitted as well as backscattered signal) to approximately five channels, similar to the experimental conditions.
In this regime, we can safely assume that electron interactions can be efficiently screened and the Landauer-B\"uttiker approach is valid.
In this two-terminal setup, the upper and lower current oscillate in-phase and several phase jumps are observed when sweeping the side gate voltage $V_s$ as imposed by the phase rigidity.  One also remarks the symmetry with respect to magnetic field as imposed by the Onsager relations. Let us note that the values for the side gate voltage are much smaller for the simulated data to induce a phase jump.
This difference is simply due to electron screening, which is not taken into account in the simulations. One can evaluate a scaling factor of about 20-30 by comparing the pinch-off voltages of the individual gates between experiment and simulations. 

\begin{figure}[htbp]
	\includegraphics[width=8cm]{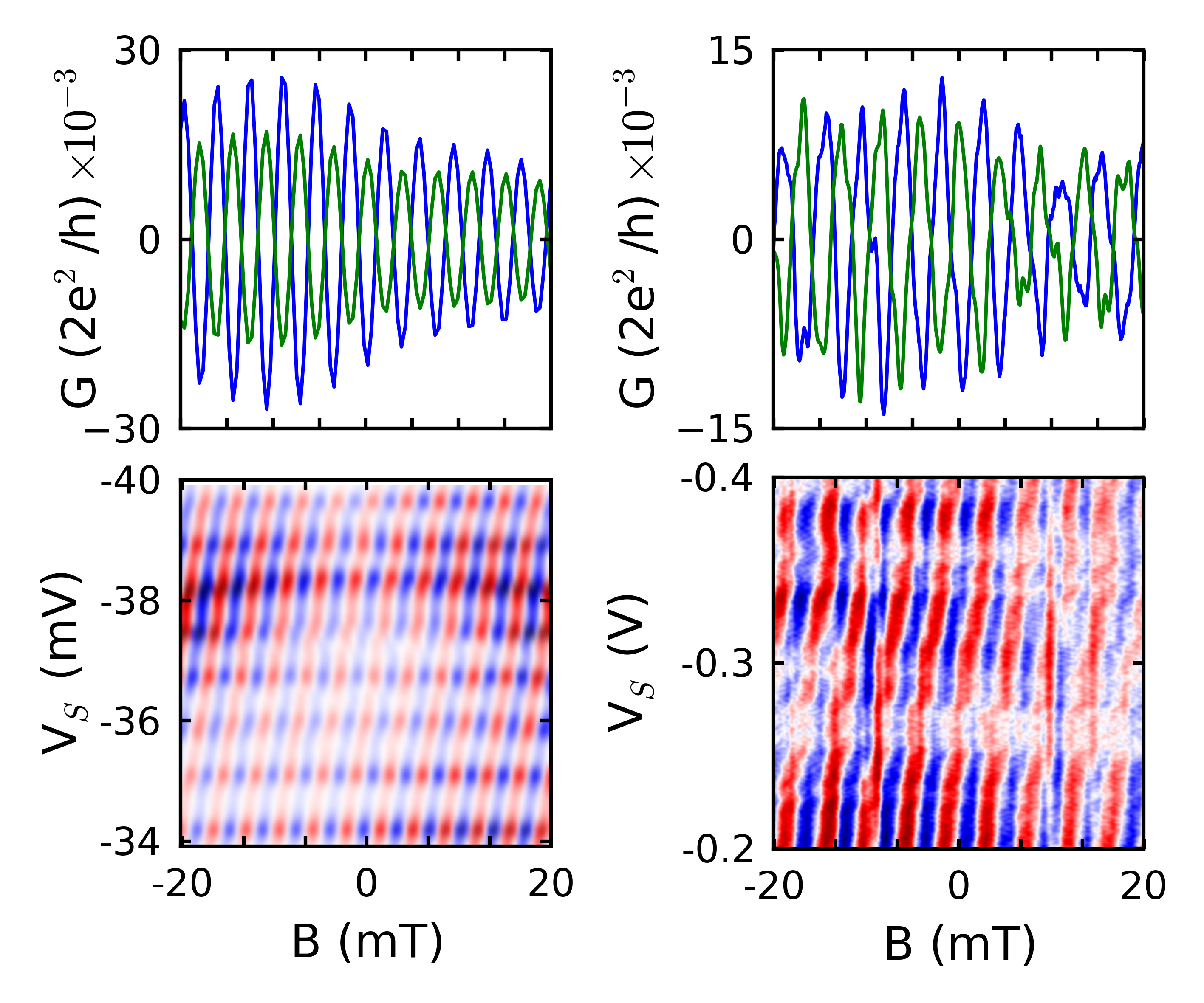}
	\caption{Magneto conductance oscillations in the crossover region between the SCR and WCR after subtraction of a smooth background: left simulations, right experimental data. Top: magneto conductance oscillations for $V_T$ = -8.3 mV. The blue (green) curve corresponds to the current in the upper (lower) contact. Bottom: 2D colour plot of the magneto conductance oscillations of the total transmitted current  (upper - lower) as a function of side gate voltage. For both data sets one observes a smooth phase shift of the AB oscillations as a function of side gate voltage $V_{s}$.}
	\label{fig:anti-phase}
\end{figure}

The more interesting regime, however, is the weak coupling regime when a finite tunnel coupling is induced by the right tunnel-coupled wire.
In this case one observes the appearance of anti-phase oscillations when imposing a finite gate voltage on the tunnel gate as shown in Fig.\ \ref{fig:anti-phase}. 
The simulated data (left panel) reproduces nicely the experimentally observed  anti-phase oscillations (right panel).
From the simulations we find that anti-phase oscillations appear even by imposing only a very small tunnel barrier. 
For the present simulations $V_T$ has been set to -8.3 mV where we are basically in a single wire regime with a very flat bottom for the electrostatic potential. 
We associate the appearance of the anti-phase oscillations to a situation where the potential change at the transition region between the AB ring and the tunnel-coupled wire is such that the symmetric modes of the AB ring can be smoothly coupled to the symmetric and anti-symmetric modes within the tunnel-coupled wire which finally leads to anti-phase oscillations. It should hence be possible to induce anti-phase oscillation in a single wire when shaping carefully the potential landscape. 
In such a "peculiar" single wire regime, one is also able to induce a smooth shift of the AB oscillations when sweeping the side gate voltage $V_S$.
This is shown in the bottom panels of Fig.\ \ref{fig:anti-phase}. Features (P1b \& P2b) are hence nicely reproduced by the simulations.
The absolute amplitude of the conductance oscillations is very sensitive to the side gate voltage and can vary between (0.01 - 0.1) $\times$ $2e^2/h$ for the investigated gate voltage scan.
Let us note that the smoothness of the phase shift is sensitive to symmetry of the gate voltages applied to the individual gates.
For instance, if we induce an asymmetry of two equivalent gates the phase shift becomes more irregular and the anti-phase oscillations are not perfectly anti-phase any more.
This is also observed in the experiments.

\begin{figure}[htbp]
	\includegraphics[width=8cm]{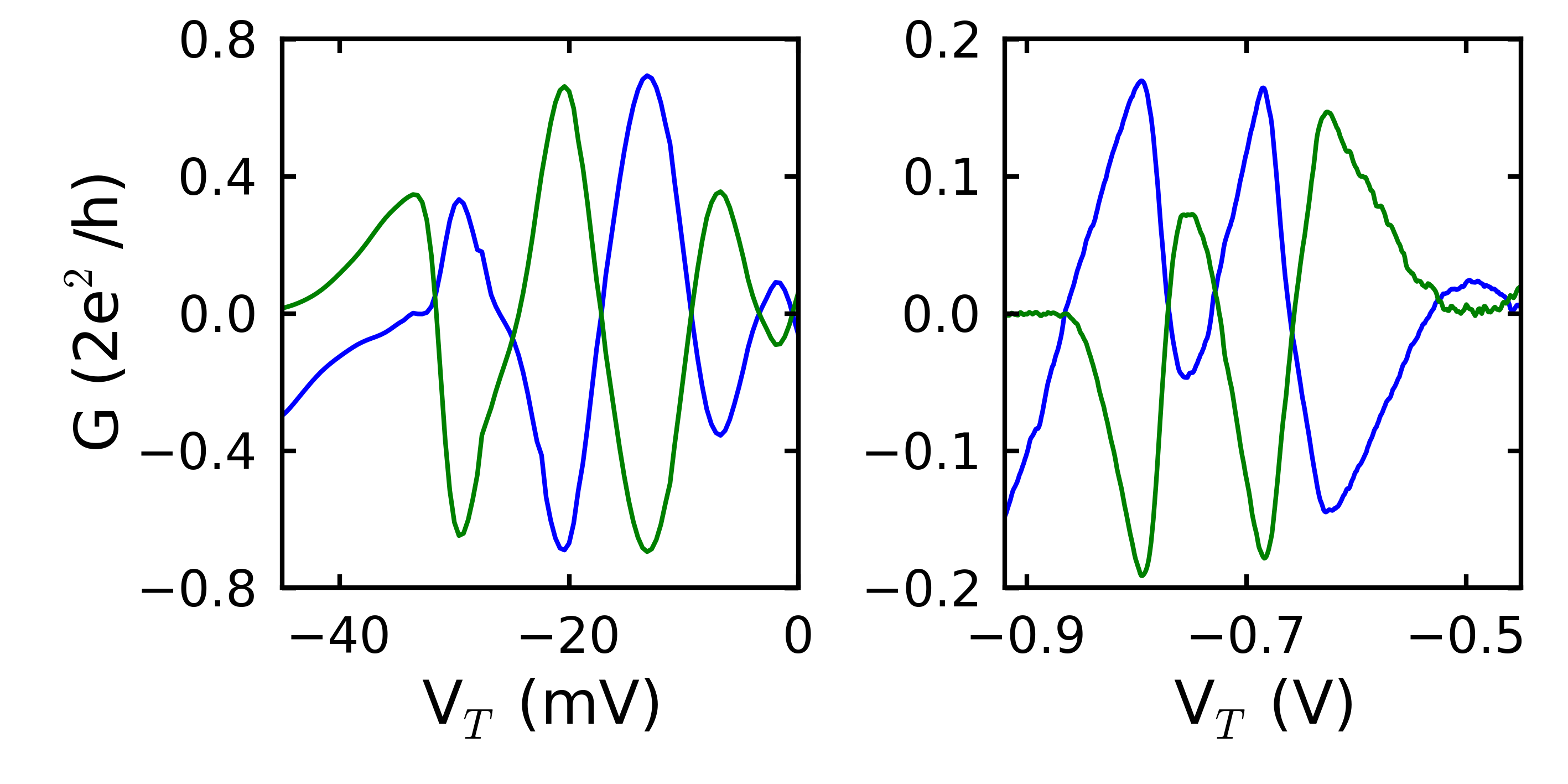}
	\caption{Conductance as a function of tunnel gate voltage after subtraction of a smooth background: left simulations, right experimental data. The blue (green) curve corresponds to the current in the upper (lower) contact. }
	\label{fig:V_T-oscillation}
\end{figure}

The most interesting observation of the experiment is certainly the conductance oscillations with respect to the tunnel gate voltage in the WCR.
This allows to partition the output current into the upper/lower channels and hence tune the tunnel-coupled wire into a beam splitter regime.
In this case the left tunnel gate is fully depleted to inject the current only into the lower branch of the AB interferometer.
In Fig.\ \ref{fig:V_T-oscillation} we show the simulated as well as the experimental data.
While at very small tunnel gate voltage (SCR) the two output currents are basically equal, we observe anti-phase oscillations for both data sets when approaching the WCR. For strongly negative gate voltage the tunnel barrier splits the tunnel-coupled wire into two independent wires.
Again the correspondence between experiment and simulation is quite remarkable. Let us note, however, that in the experiment for a gate voltage regime below the 2D pinch-off ($V_T \approx$ -0.3 to -0.5 V) the oscillations are suppressed. 
This is most probably due to density alignment of one of the subbands caused by electron-electron interactions \cite{energy,michi_nnano_12}, which sets the corresponding channel into off-resonance. Such influences of the Coulomb interaction is not taken into account in our model. 

\begin{figure}[htbp]
	\includegraphics[width=8cm]{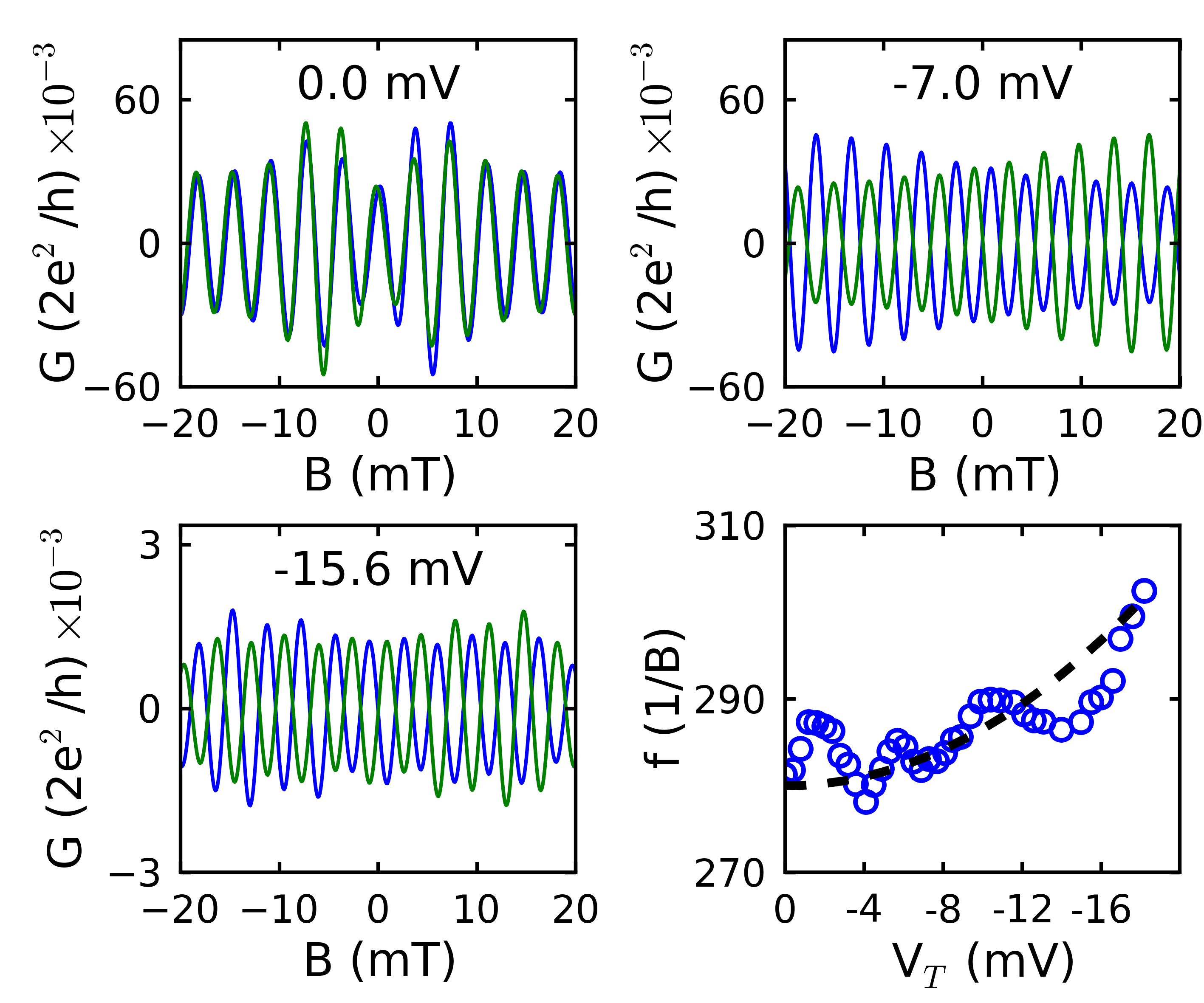}
	\caption{Simulations of the evolution of the AB conductance oscillations for different values of the tunnel gate voltage $V_T$. A smooth background has been subtracted from the data. Bottom right panel:  frequency  of the magneto oscillations (obtained by FFT of the other panels) as a function of $V_T$.}
	\label{fig:surface}
\end{figure}

Analysing the magneto conductance oscillations as a function of tunnel gate voltage $V_T$, we observe in the numerical simulations a change in the magneto conductance oscillation frequency when passing from the SCR to the WCR as displayed in Fig.\ \ref{fig:surface}. For the SCR ($V_T$ = 0V) we observe in-phase oscillations as expected and the oscillation period corresponds simply to the surface area enclosed by the AB loop. 
When increasing the tunnel barrier height (decrease of tunnel gate voltage) one clearly observes an increase of the number of periods for a given magnetic field scan.
This implies that the effective AB surface area increases. We emphasize this by taking the fast Fourier transform (FFT) of the magneto oscillations and by plotting the FFT peak position as a function of the tunnel gate voltage (bottom panel).
One clearly sees an increase of the Fourier peak as a function of tunnel gate voltage as indicated by the dashed line in Fig.\ \ref{fig:surface}.

To explain this observation, the electron has to pick up an AB phase over a significant distance in the tunnel-coupled wire. We associate this effect to the presence of the magnetic field which displaces the wave function with respect to the center of the tunnel-coupled wire due to the Lorentz force.
As a consequence, the electrons will acquire an additional Aharonov-Bohm phase which is proportional to $\frac{\hbar}{e} \frac{\partial \Delta k }{ \partial B} L$, in agreement with the semi analytical results of Fig.\ \ref{fig:bonding+antibonding} (bottom panel).
This can be interpreted as a surface area increase and explains the observed change in the magneto oscillation period in the simulations when going from the SCR to the WCR. 
Note however, that the Lorentz force makes the symmetric and anti-symmetric states more localized in either of the two wires (see appendix) and induces an imbalance of the coupling of these states to the upper and lower wire. As a consequence, the visibility decreases with increasing surface area. In addition, increasing the length of the tunnel-coupled wire enhances this surface area increase almost linearly. These effects could readily be tested with the experimental set-up of ref. \cite{michi_nnano_12}. 
Naturally, it would also be interesting to implement electron interactions into the numerical simulations \cite{sukhorukov_09} to allow for better quantitative agreement between theory and experiment as well as the possibility to study other effects such as decoherence \cite{lindelof_01,niimi_09,niimi_10,preden_08,sukhorukov_09}.

 \section{\bf Conclusion}

We have presented a minimum scattering theory as well as realistic simulations of an Aharonov-Bohm interferometer connected to two tunnel-coupled wires, a solid state implementation of a flying qubit.
While our simplified model could account for most experimental observations by assuming suppression of backscattered induced loop trajectories, our numerical simulations of the actual experimental system with realistic parameters can reproduce the majority of the experimentally observed features as well as suppression of  multiple loops in the AB ring. These simulations are important in particular for the understanding of rather subtle, geometry related aspects.
The good agreement between experiment and theory shows, that the physics of the flying qubit is well described within the Landauer-B\"uttiker scattering formalism. In addition to the interpretation of the experiments, the sort of simulations performed with Kwant could be a useful tool for quantum device design and signal optimisation.


 \section{\bf acknowledgements}
C.B. would like to thank M. B\"uttiker, L. Glazman, T. Jonckheere, T. Kato, T. Martin, C. Texier and  R. Withney for useful discussions.
T.B. acknowledges financial support from the Nanoscience Foundation Grenoble. S. Takada acknowledges support from JSPS Research Fellowships for Young Scientists. M.Y. acknowledges financial support by Grant-in-Aid for Young Scientists A (no. 23684019). S. Tarucha acknowledges financial support by MEXT KAKENHI "Quantum Cybernetics", MEXT project for Developing Innovation Systems, and JST Strategic International Cooperative Program. X.W. acknowledges financial support from the ERC grant MesoQMC. C.B. acknowledges partial financial support from the French National Agency (ANR) in the frame of its program BLANC FLYELEC project no anr-12-BS10-001 and from CNRS (DREI) - JSPS (PRC0677).

 \section{\bf appendix}

\begin{figure}[htbp!]
	\includegraphics[width=8cm]{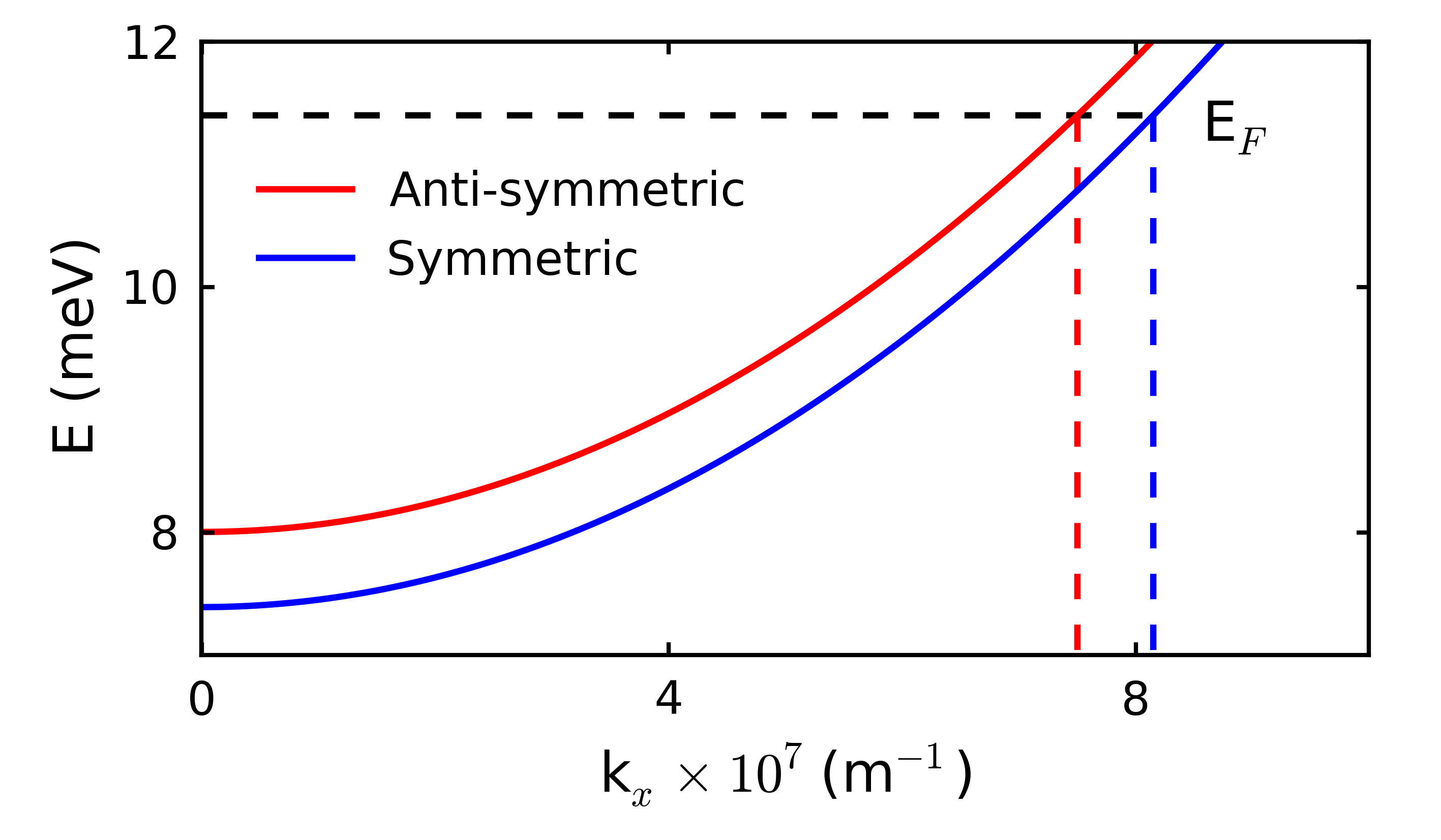}
	\caption{Dispersion relations for the symmetric (anti-symmetric) state in blue (red). The wave vectors in propagation direction have been extracted at the Fermi energy as indicated by the dashed lines.}
	\label{fig:appendix01}
\end{figure}

{\bf Dispersion relation: }For each set of experimental parameters, e.g. tunnel gate voltage and magnetic field, we calculate the dispersion relations for the symmetric and anti-symmetric state as shown in Fig. \ref{fig:appendix01}  for  $V_T = 0\,$V and $B =0\,$T. From this we can extract the wave vectors in propagation direction $\hat x$ for each mode at the Fermi energy E$_F$ and hence $\Delta k$. By taking $k_F^2=k_y^2+k_x^2$ we can also compute the eigenenergies of these two states due to confinement. 

{\bf Magnetic field dependence:} At zero magnetic field, the symmetric and anti-symmetric state are degenerate at high tunnel gate voltage. 
However, when a magnetic field is applied, the situation becomes rather subtle. 
To understand the underlying physics, we consider the two-dimensional Schr\"odinger equation (\ref{eq:sql}).

The vector potential can be expressed within the Landau-gauge
\begin{equation}
\vec{A}=-By\hat{e_x},
\end{equation}
which leads after separation of variables to
\begin{equation}
\begin{split}
[\hbar^2k_x^2+2e\hbar Byk_x+e^2B^2y^2-\hbar^2\Delta+V(y)]\Psi(y)\\
=E\Psi(y)2m.
\end{split}
\end{equation} 

\begin{figure}[htbp!]
	\includegraphics[width=8cm]{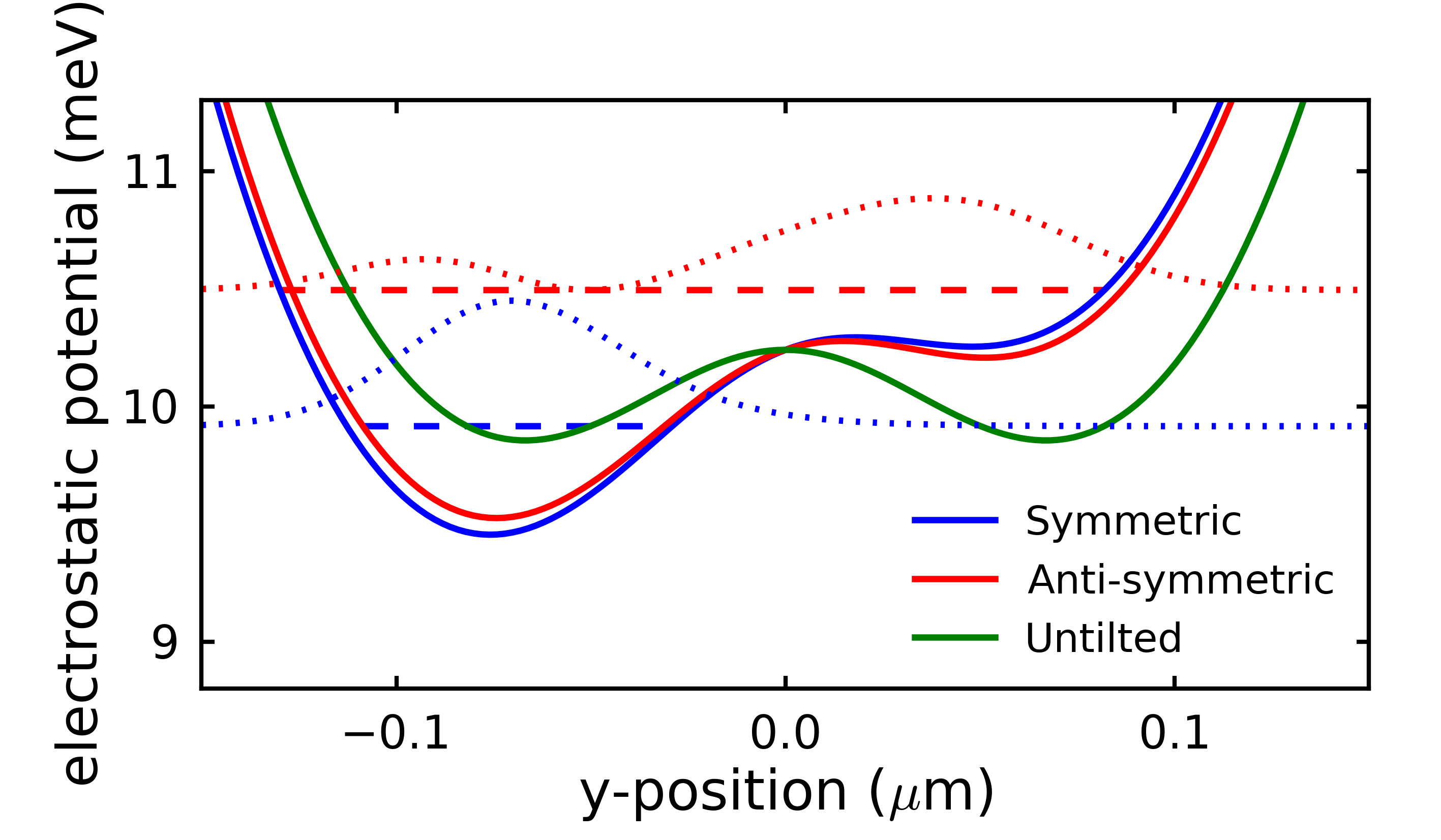}
	\caption{Effective confinement potential under magnetic field of 60 mT. Blue (red) solid line corresponds to the symmetric (anti-symmetric) state. The dashed lines show their respective eigenenergies. Green solid line shows the confinement potential for $B = 0\ {\rm T}$ for comparison. The blue (red) dotted curves correspond to the symmetric (anti-symmetric) wave function under magnetic field.}
	\label{fig:appendix02}
\end{figure}

We can now identify three spatially dependent terms: V(y) denotes the electrostatic potential in the tunnel-coupled wire created by the surface gates. The quadratic term increases the parabolic confinement symmetrically, whereas the second term induces a tilt in the potential landscape which is linear in y, B and $k_x$. This leads to a energy difference of the symmetric and anti-symmetric state and hence to a finite $\Delta k$ as depicted in Fig. \ref{fig:appendix02}.

\newpage

\bibliography{flying-qubit}{}

\begin{thebibliography}{39}
\bibitem[{ref(.)}]{ref-quant}
\bibinfo{journal}{this can be done directly with Kwant using the solver:
  kwant.physics.Band}.
\end{thebibliography}

\end{document}